\DeclareSymbolFont{matha}{OML}{txmi}{m}{it}
\DeclareMathSymbol{\varv}{\mathord}{matha}{118}

\documentclass[twoside,twocolumn,9pt]{article}
\usepackage[super,sort&compress,comma]{natbib} 
\usepackage[version=3]{mhchem}
\usepackage[left=1.5cm, right=1.5cm, top=1.785cm, bottom=2.0cm]{geometry}
\usepackage{balance}
\usepackage{mathptmx}
\usepackage{sectsty} 
\usepackage{graphicx} 
\usepackage{lastpage}
\usepackage[format=plain,justification=justified,singlelinecheck=false,font={stretch=1.125,small,sf},labelfont=bf,labelsep=space]{caption}
\usepackage{float}
\usepackage{fancyhdr}
\usepackage{fnpos}
\usepackage[english]{babel}
\addto{\captionsenglish}{%
  
}
\usepackage{array}
\usepackage{droidsans}
\usepackage{charter}
\usepackage[T1]{fontenc}
\usepackage[usenames,dvipsnames]{xcolor}
\usepackage{setspace}
\usepackage[compact]{titlesec}
\usepackage{hyperref}

\usepackage{epstopdf}

\definecolor{cream}{RGB}{222,217,201}

\usepackage{placeins}
\usepackage[normalem]{ulem}

    \newcommand{\blu}[1]{{\color{black}{#1}}}

\newcommand*{\citen}[1]{
  \begingroup
    \romannumeral-`\x 
    \setcitestyle{numbers}%
    \cite{#1}%
  \endgroup   
}

\begin{document}

\pagestyle{fancy}
\thispagestyle{plain}
\fancypagestyle{plain}{
\renewcommand{\headrulewidth}{0pt}
}

\makeFNbottom
\makeatletter
\renewcommand\LARGE{\@setfontsize\LARGE{15pt}{17}}
\renewcommand\Large{\@setfontsize\Large{12pt}{14}}
\renewcommand\large{\@setfontsize\large{10pt}{12}}
\renewcommand\footnotesize{\@setfontsize\footnotesize{7pt}{10}}
\makeatother

\renewcommand{\thefootnote}{\fnsymbol{footnote}}
\renewcommand\footnoterule{\vspace*{1pt}%
\color{cream}\hrule width 3.5in height 0.4pt \color{black}\vspace*{5pt}} 
\setcounter{secnumdepth}{5}

\makeatletter 
\renewcommand\@biblabel[1]{#1}            
\renewcommand\@makefntext[1]%
{\noindent\makebox[0pt][r]{\@thefnmark\,}#1}
\makeatother 
\renewcommand{\figurename}{\small{Fig.}~}
\sectionfont{\sffamily\Large}
\subsectionfont{\normalsize}
\subsubsectionfont{\bf}
\setstretch{1.125} 
\setlength{\skip\footins}{0.8cm}
\setlength{\footnotesep}{0.25cm}
\setlength{\jot}{10pt}
\titlespacing*{\section}{0pt}{4pt}{4pt}
\titlespacing*{\subsection}{0pt}{15pt}{1pt}

\fancyfoot{}
\fancyfoot[LO,RE]{\vspace{-7.1pt}\includegraphics[height=9pt]{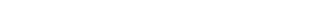}}
\fancyfoot[CO]{\vspace{-7.1pt}\hspace{13.2cm}\includegraphics{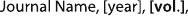}}
\fancyfoot[CE]{\vspace{-7.2pt}\hspace{-14.2cm}\includegraphics{RF}}
\fancyfoot[RO]{\footnotesize{\sffamily{1--\pageref{LastPage} ~\textbar  \hspace{2pt}\thepage}}}
\fancyfoot[LE]{\footnotesize{\sffamily{\thepage~\textbar\hspace{3.45cm} 1--\pageref{LastPage}}}}
\fancyhead{}
\renewcommand{\headrulewidth}{0pt} 
\renewcommand{\footrulewidth}{0pt}
\setlength{\arrayrulewidth}{1pt}
\setlength{\columnsep}{6.5mm}
\setlength\bibsep{1pt}

\makeatletter 
\newlength{\figrulesep} 
\setlength{\figrulesep}{0.5\textfloatsep} 

\newcommand{\topfigrule}{\vspace*{-1pt}%
\noindent{\color{cream}\rule[-\figrulesep]{\columnwidth}{1.5pt}} }

\newcommand{\botfigrule}{\vspace*{-2pt}%
\noindent{\color{cream}\rule[\figrulesep]{\columnwidth}{1.5pt}} }

\newcommand{\dblfigrule}{\vspace*{-1pt}%
\noindent{\color{cream}\rule[-\figrulesep]{\textwidth}{1.5pt}} }

\makeatother

\twocolumn[
  \begin{@twocolumnfalse}
{\includegraphics[height=30pt]{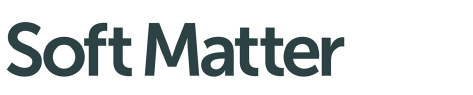}\hfill\raisebox{0pt}[0pt][0pt]{\includegraphics[height=55pt]{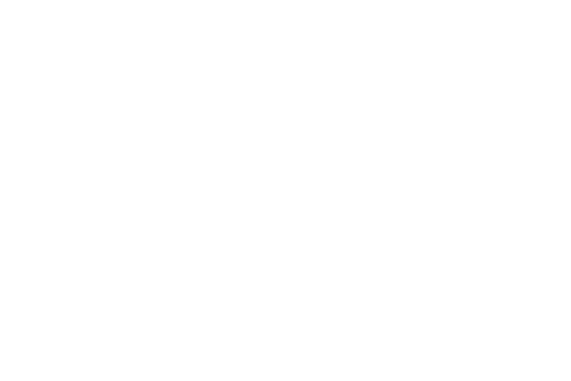}}\\[1ex]
\includegraphics[width=18.5cm]{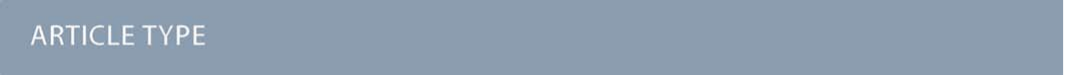}}\par
\vspace{1em}
\sffamily
\begin{tabular}{m{4.5cm} p{13.5cm} }

\includegraphics{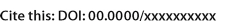} & \noindent\LARGE{\textbf{Molecular Dynamics Simulation of the Coalescence of Surfactant-Laden Droplets$^\dag$}} \\
\vspace{0.3cm} & \vspace{0.3cm} \\

 & \noindent\large{Soheil Arbabi,\textit{$^{a}$} Piotr Deuar,\textit{$^{a}$} Mateusz Denys,\textit{$^{a}$} Rachid Bennacer,\textit{$^{b}$} Zhizhao Che,\textit{$^{c}$} and Panagiotis E. Theodorakis$^{\ast}$\textit{$^{a}$}} \\

\includegraphics{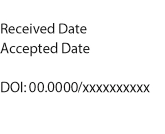} & \noindent\normalsize{We investigate the
coalescence of surfactant-laden water droplets by using
several different surfactant types and a wide range of
concentrations by means of a coarse-grained model obtained by
the statistical associating fluid theory. 
Our results demonstrate in detail a universal
mass transport mechanism of surfactant
across many concentrations and several surfactant types during the process. 
Coalescence initiation is seen to occur via a single 
pinch due to aggregation of surface surfactant, 
and its remnants tend to become engulfed in part inside 
the forming bridge. Across the board we confirm 
the existence of an initial thermal regime 
with constant bridge width followed by a 
later inertial regime with bridge width 
scaling roughly as the square root of time, 
but see no evidence of an intermediate viscous regime. 
Coalescence becomes slower as surfactant concentration grows, 
and we see evidence of the appearance of a further 
slowdown of a different nature for several times 
the critical concentration. We anticipate that our results
provide further insights in the mechanisms of coalescence
of surfactant-laden droplets. 
}

\end{tabular}

 \end{@twocolumnfalse} \vspace{0.6cm}

  ]

\renewcommand*\rmdefault{bch}\normalfont\upshape
\rmfamily
\section*{}
\vspace{-1cm}


\footnotetext{\textit{$^{a}$~Institute of Physics, Polish Academy of Sciences, Al. Lotnik\'ow 32/46, 02-668 Warsaw, Poland. 
}}
\footnotetext{\textit{$^{b}$~Université Paris-Saclay, CentraleSupélec, ENS Paris-Saclay, CNRS, LMPS - Laboratoire de Mécanique Paris-Saclay, 91190, Gif-sur-Yvette, France}}
\footnotetext{\textit{$^{c}$~State Key Laboratory of Engines, Tianjin University, 300350 Tianjin, China. }}
\footnotetext{\dag~Electronic Supplementary Information (ESI) available: [details of any supplementary information available should be included here]. See DOI: 10.1039/cXsm00000x/}



\section{Introduction}
\label{intro}
While ubiquitous in nature, droplet coalescence is also  
an important process in various industrial 
applications, where the rate of coalescence can 
determine their performance. 
For example, in the context of bio-related
microfluidic technologies, slowing down
coalescence in bio-particle encapsulation
on lab-on-chip devices is often desirable,
and can be achieved by using various additives,
such as surfactants.\cite{feng2015advances,baret2012surfactants}
In contrast, speeding up the rate of coalescence
by adding surfactant could be advantageous in applications, 
such as coatings\cite{ristenpart2006coalescence} and 
superspreading.\cite{Theodorakis2014,Theodorakis2019molecular}

Experimental, theoretical, and numerical studies
of coalescing droplets have thus far mainly focused on
cases without additives, \textit{e.g.} water or polymer
droplets.\cite{Paulsen2014,yoon2007coalescence,khodabocus2018scaling,perumanath2019droplet,eggers1999coalescence,aarts2005hydrodynamics,sprittles2012coalescence,Dudek2020,Rahman2019,Berry2017,Somwanshi2018,Kirar2020,Bayani2018,Brik2021,Anthony2020,Kern2022,Heinen2022,Geri2017,Abouelsoud2021,Arbabi2023,Dekker2022,Calvo2019,Sivasankar2022,Otazo2019,Vannozzi2019}
From the point of view of numerical simulations, these have
by and large provided descriptions of the macroscopic and dynamic
properties of coalescence,
\cite{yeo2003film,hu2000drop,yoon2007coalescence,mansouri2014numerical,khodabocus2018scaling,Anthony2020,Kern2022,Heinen2022}
but they generally continue to suffer from inadequate resolution at the pinching
point between droplets at the initial stage of 
coalescence, despite progress in this area.\cite{sprittles2012coalescence}
Moreover, a detailed molecular-level
description of the mass transport mechanism of surfactant
is beyond the reach of any continuum model.\cite{eggers1999coalescence,duchemin2003inviscid}

Various attempts have been made to tackle this challenge.
For example, numerical models based on advection--diffusion 
equations and chemical kinetic fluxes attempt to
incorporate the mass transport of surfactant into the
equations, \cite{karapetsas2011}
but these are only as good as the assumptions put 
into the model, while a detailed molecular description still remains out of reach for continuum simulation.
When it comes to experimental techniques,\cite{nowak2016effect, narayan2020zooming, nowak2017bulk, duchemin2003inviscid, leal2004flow,Politova2017,Dong2017,Dong2019,Weheliye2017,Kovalchuk2019}
high-speed imaging 
and particle-image velocimetry
have been applied in investigating coalescence
of surfactant-laden droplets, and have
mainly focused on the macroscopic description
of this phenomenon, much as in the case of numerical
simulation. Due to device limitations,
capturing with high resolution the early stages of coalescence 
during experiments poses a major challenge.
Both experiments and continuum modelling are unable to
provide a detailed description of the mass transport mechanism
at the molecular level.
In the case of molecular simulations, 
all-atom molecular dynamics (MD) has explained 
different microscopic aspects of coalescence,
such as the role of the capillary waves on droplets' 
surface, but these have only been in the context of pure 
water droplets,\cite{perumanath2019droplet} 
except for one recent study which considered a few cases.\cite{arbabi2023b}

On the whole, though, the effect of surfactant in droplet coalescence has largely
remained unexplored, 
\blu{with implications actually reaching beyond this,
for example, in the context of surfactant-laden coalescence between micelle
and bilayer\cite{Li2008}}. The initial work\cite{arbabi2023b} 
suggests that it plays
an important role, \textit{e.g.} affecting the coalescence 
rate and dampening internal droplet dynamics.
Reduction of the surface tension at
the liquid--gas (LG) interface is usually expected, but
the type of surfactant (\textit{e.g.,} chemical
groups and molecular architecture) and its
concentration are also anticipated
to differently affect the coalescence process. 
Differences are expected even when the concentration is higher than
the critical aggregation concentration (CAC)
and surface tension at the LG interface is expected to remain 
constant.\cite{Ritacco2010,Ritacco2010b,Ivanova2010,Ivanova2012,Ivanova2009}
On that account, unveiling the details of this phenomenon 
from a molecular-level perspective is key
for fundamentally understanding the underlying
mechanisms that can lead to tailor-made
surfactant designs for applications.

Here we report on a more detailed and robust study 
of droplet coalescence using the methods of Ref.~\citen{arbabi2023b},
and a broader range of surfactants and their concentrations.
We employed a high-fidelity coarse-grained
force-field enabling us to simulate with MD 
the coalescence of surfactant-laden droplets.
We discuss the details
of coalescence at each stage of the process
by comparing systems with different surfactant
type and concentration. Our data indicate
that the underlying mechanisms of coalescence are
consistent between the different cases,
which point to universal features of surfactant transport and 
droplet dynamics for the coalescence of water droplets with surfactant. 
Differences in the behaviour of different surfactants are also highlighted
when relevant. The dynamic and static characteristics
are unveiled by discussing both macroscopic- 
and molecular-level quantities,
such as the bridge growth dynamics and the velocity of 
droplet approach, as well as describing the
molecular transport of surfactant within the liquid phase.
Thus, we anticipate that the present 
study lays a comprehensive account of 
the coalescence of surfactant-laden droplets.

In the following section, we provide background
information for droplet coalescence.
Then, Section~\ref{model} gives details
on our model and methodology, while Section~\ref{results}
presents and discusses the
results of the coalescence simulations.
Section~\ref{conclusions}, draws some broader conclusions.

\begin{figure}[bt!]
\includegraphics[width=\columnwidth]{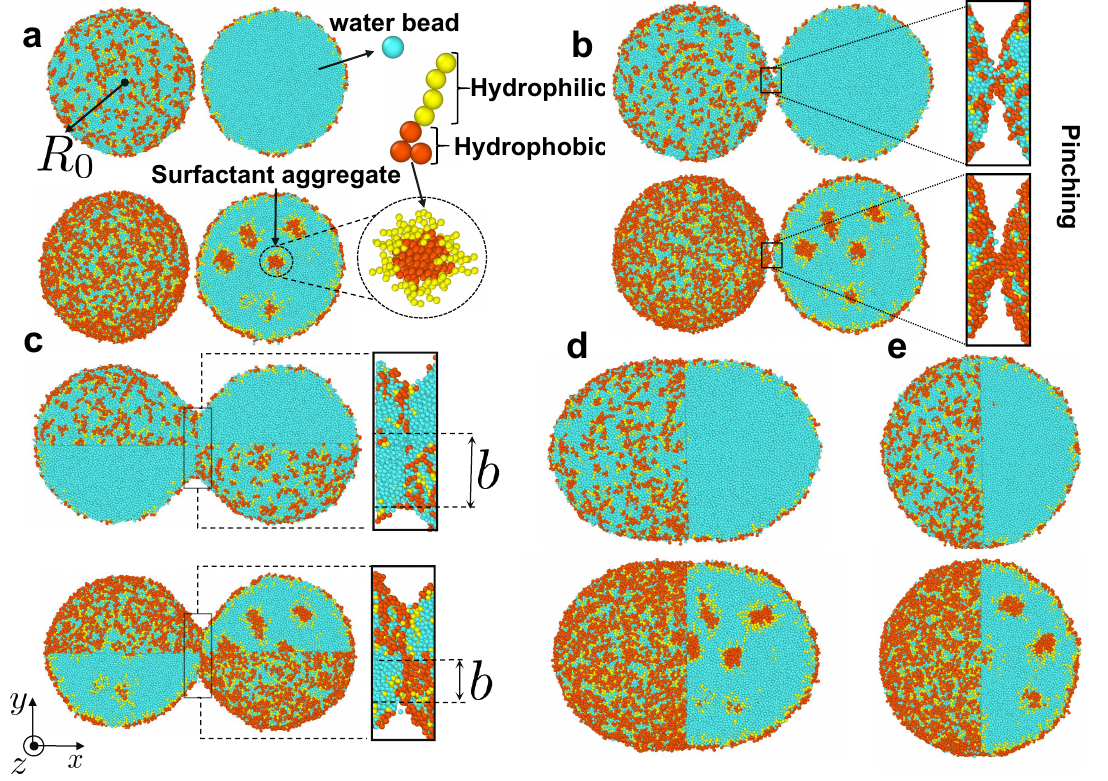}
\caption{\label{fig:1} Coalescence of droplets
with C10E4 surfactant at low concentration 
(6.25~wt\%, top of each panel) below $\text{CAC}=7.5$~wt\% 
and  concentrations above CAC (24.18~wt\%, bottom of 
each panel). Snapshots show the
initial approach of the droplets (a), 
their pinching and initial formation of the bridge (b), see
also ESI for a movie showing the pinching of
the droplets),
configurations with a partially (c) and fully (d) 
developed bridge, and the final equilibrium
state (e). External or cross-section views are shown
to highlight the bulk, surface, and bridge structure
of the droplets. Magnified views of a micelle, a surfactant molecule,
and a water bead are shown in panel (a). Water vapour 
surrounding the droplets is not shown for the sake of clarity.
}
\end{figure}

\section{Background}
\label{background}
A macroscopic description of the droplet coalescence
can generally be divided into three stages 
(see Fig.~\ref{fig:1}). The 
approach of the droplets, which leads to the 
initiation of the phenomenon (pinching) as a result of 
the inter-molecular interactions; the subsequent
growth of the bridge formed between the droplets;
and the final stage towards equilibrium, which
manifests itself by the formation of a single spherical
droplet.
The system is driven towards the equilibrium state as it
seeks to minimise the free-energy associated with the
surface tension at the droplet boundary. 
Hence, on the coarsest level, the formation
of larger droplets is energetically favourable. 
In particular, for a droplet with $N$ beads,
the droplet radius behaves as $R_0\propto N^{1/3}$ (Fig.~\ref{fig:1}a) 
since the volume scales as $N$,
while surface tension force as $\propto N^{2/3}$. 

The rate of bridge growth after the initial contact
is key for characterising
the coalescence process and can generally be
described from the point of view of 
fluid dynamics by two successive regimes,
namely the viscous regime (VR) and the inertial 
regime (IR).\cite{Paulsen2014,paulsen2013approach}
Moreover, recent MD simulations have uncovered the existence
of a third, thermal regime (TR) during the droplet pinching at the
very initial stage of the coalescence 
process,\cite{perumanath2019droplet,arbabi2023b}
which will be discussed later in more detail.
The characteristic velocity in the VR can
be defined as $\varv_{\rm v}=\gamma/\eta$,
where $\gamma$ is the surface tension 
and $\eta$ the viscosity,
which suggests that the capillary number
${\rm Ca}=\eta \varv_{\rm v} / \gamma \sim 1$, 
with the characteristic time scale for droplet-sized effects being 
$t_{\rm v}=R_0 \eta / \gamma$ \cite{eggers1999coalescence}, 
with $R_0$ the droplet radius as in Fig.~\ref{fig:1}.
As coalescence proceeds to the subsequent
IR, the bridge flow
can be characterised by the Weber number, namely
${\rm We} = \rho \varv_{\rm i}^2 R_0 / \gamma \sim 1$
indicating the limit that inertia effects will take over
the surface tension effects, while $\rho$
is the density of the fluid.
The bridge velocity is found to scale as
$\varv_{i} \sim\sqrt{\gamma / \rho R_0}$,
and thus the characteristic time scale is 
$t_{\rm i} = \sqrt{\rho R_0^3 / \gamma}$.
For many purposes, especially in the VR,  
the more relevant length scale is the radius of the bridge, $b$ 
(Fig.~\ref{fig:1}), and the corresponding
Reynolds number can be defined as Re\,$=\rho \varv 
b / \eta$, which in the VR is $\rho \gamma b/ \eta^2$.
On the one hand, since the bridge radius is very
small in the initial stage, the viscous forces 
are dominant regardless of the values of
$\gamma$ and $\eta$ and Re\,$\ll1$. 
On the other hand, Re\,$\ge 1$ reflects the IR.\cite{gross2013viscous}
Hence, it can be taken that the crossover
between the viscous and the inertial regimes
would take place for ${\rm Ca} \sim 1$ or ${\rm Re} \sim 1$.
Taking the IR expression $\varv_{\rm i} \sim \sqrt{\gamma/b\rho}$ 
using bridge size, we obtain an estimate of the
crossover bridge-radius, $b_{\rm c} = \eta^2 / \rho \gamma$,
and a characteristic time $t_{\rm c} = \eta^3 / \rho \gamma^2$,
which for water droplets would roughly correspond to
$b_{\rm c} \sim 15$\,nm and $t_{\rm c} = 0.1$\,ns, highlighting the
fast time scales of coalescence.\cite{aarts2005hydrodynamics}

In the VR, where inter-molecular 
forces are playing a dominant role in pulling
the droplets together, a linear scaling has been
proposed for the growth of the bridge radius 
with time, \textit{i.e.} $b\propto t$, as well as
logarithmic corrections, $b \propto t \ln t$.\cite{duchemin2003inviscid,eggers1999coalescence}
In the case of the IR, a power-law
scaling law has been suggested for the bridge, 
namely $b\propto \sqrt{t}$.\cite{duchemin2003inviscid,eggers1999coalescence}
Experimental studies on the coalescence of
water droplets are consistent with this 
and have shown that 0.7~ns after the first
contact of the droplets, drag forces give way to
the inertial ones and the bridge radius has been
reported to scale with time as 
$b \propto (R_{0}\gamma/\rho)^{1/4}t^{1/2}$ where 
 $R_0$ is the initial droplet radius. \cite{eggers1999coalescence,aarts2005hydrodynamics,thoroddsen2005coalescence,gross2013viscous,sprittles2012coalescence}
Other works have proposed scaling regimes
that depend on the ratio of characteristic scales
to the viscous length scale $l_v = \eta^2/\rho\gamma$.\cite{paulsen2012inexorable,paulsen2013approach}
Moreover, it is argued that the inertia
of the droplets cannot be neglected
at the initial stage of coalescence. Then, the initial stage would
be better described as inertially limited viscous (ILV) 
regime and a linear scaling with time for the bridge radius
has been proposed, which,
according to numerical simulations,
is only realised when the coalescing drops are 
initially separated by a finite distance.\cite{Anthony2020}
In the case of miscible and immiscible droplets, 
a similar viscous dominated regime has been
suggested, but immiscible droplets seem to
develop a bridge slower.\cite{xu2022bridge}

All-atom molecular dynamics simulations 
of pure two-dimensional (cylindrical) water droplets have found that multiple 
precursor bridges develop via thermal fluctuations at the droplet's surface,
which initially connect the droplets and then grow, thus identifying
a thermal regime at the onset of coalescence.\cite{perumanath2019droplet}
After a certain threshold, when the radius of the growing bridge
becomes larger than a thermal length scale, 
$l_T  \approx {\left ( k_BT/\gamma \right) }^{1/4} R_{0}^{/1/2}$,
the hydrodynamic regime is recovered,
continuum models can describe the process, 
and one expects VR, IR, or ILV scaling.
$k_B$ is Boltzmann's constant and $T$ the temperature.  
Since surface tension appears in the denominator, 
the addition of surfactant, which lowers the surface tension,
is expected to lead to the increase of the
thermal length, $l_T$, towards an upper limit
that is attained above the CAC. Moreover, these all-atom MD
simulations have found that, in the TR, the bridge radius
expands linearly in time with a velocity much faster
than the viscous--capillary regime due to the thermal,
molecular `jumps' at the droplets' surface where multiple
bridges are formed.\cite{perumanath2019droplet}
The initial thermal regime has also been recently observed
in the case of coarse-grained models for pure and
surfactant-laden three-dimensional (spherical) droplets, 
without the formation of multiple thermal bridges
but a single pinching point instead.\cite{arbabi2023b}

The above considerations indicate that coalescence
is still under intensive debate even in the case
of pure water droplets. Moreover, only a few
studies have dealt with droplet coalescence
in the presence of surfactant.\cite{Politova2017,Weheliye2017,Soligo2019,Dong2017,Dong2019,Kovalchuk2019,Botti2022,Amores2021,Kasmaee2018,nowak2017bulk,nowak2016effect,jaensson2018tensiometry,narayan2020zooming,ivanov1999flocculation,tcholakova2004role,langevin2019coalescence,Velev1993,Suja2018} 
While it is generally expected that surfactant
would decrease the surface tension of the droplets
and a delay in the process would be potentially 
forecast,\cite{leal2004flow,jaensson2018tensiometry}
comprehensive understanding is currently lacking, 
which calls for a systematic investigation of different
surfactants for a range of concentrations.
It has been experimentally shown that
the presence of surfactant would locally reduce
the surface tension on the droplet surface and a
nonuniform surfactant concentration would lead to 
surface tension gradients that would eventually
lead to surface flow (Marangoni flow) and a
rearrangement of the surfactant molecules to
counteract the gradient would
delay the coalescence process.\cite{nowak2016effect} 
One might also attempt to see whether the diffusion and
adsorption of surfactant at the droplets' LG interface
would further affect the coalescence process, 
for example, by influencing the flow field of the water molecules,\cite{arbabi2023b} 
especially when surfactant concentration is
above the CAC and surfactant aggregates are present in large
amounts within the liquid bulk.
MD simulations based on a high-fidelity coarse-grained force-field 
are capable of addressing these and other questions.

\section{Model and Methods}
\label{model}
An important motivation for choosing
the force-field for the problem at hand is the ability
to simulate relatively large droplets that could allow
for the investigation of the surfactant mass transport
mechanism with MD. Based on the fact that coarse-grained force-fields
would be a natural choice in this case, 
and our success in previous investigations on the mass transport
mechanism of surfactant in the context of the superspreading phenomenon,
\cite{Theodorakis2015modelling,Theodorakis2019molecular,Theodorakis2015Langmuir,Theodorakis2014,Theodorakis2019} 
we have embarked on carrying out our studies here
by using MD simulations based on
the SAFT (statistical associating fluid theory)
force-field.\cite{chapman1989saft,muller2001molecular,Avendano2011,Avendano2013,sergi2012coarse,muller2014force}
More specifically, a force-field based on the SAFT-$\gamma$ Mie
theory\cite{Lafitte2013} is used, which
can accurately reproduce relevant key properties
of water--surfactant systems, such as their phase behaviour
and surface tension.\cite{Theodorakis2015modelling, Theodorakis2019molecular, lobanova2014development, lobanova2016saft, morgado2016saft}

In the case of the SAFT force-field, interactions between 
different coarse-grained (CG) beads within a distance
smaller than $r_{\rm c}$ are described
via the Mie potential, 
which is mathematically expressed as
\begin{equation}
\label{equation_mie}
    U(r_{\rm ij}) = C\epsilon_{\rm ij} \left[ \left({\frac{\sigma_{\rm ij}}{r_{\rm ij}}}\right)^{\lambda_{\rm ij}^{\rm r}} - \left({\frac{\sigma_{\rm ij}}{r_{\rm ij}}}\right)^{\lambda_{\rm ij}^{\rm a}}\right],
     r_{\rm ij} \leq r_{\rm c},
    \end{equation}
where
\begin{equation}
    C = \left(\frac{\lambda_{\rm ij}^{\rm r}}{\lambda_{\rm ij}^{\rm r} - \lambda_{\rm ij}^{\rm a}}\right){\left( \frac{\lambda_{\rm ij}^{\rm r}}{\lambda_{\rm ij}^{\rm a}}\right)}^{\frac{\lambda_{\rm ij}^{\rm a}}{\lambda_{\rm ij}^{\rm r} - \lambda_{\rm ij}^{\rm a}}}.
\end{equation}
i and j are the bead types, $\sigma_{\rm ij}$ indicates the 
effective bead size,
and $\epsilon_{\rm ij}$ is the interaction strength
between any beads of type i and j. 
$\lambda_{\rm ij}^a=6$ and $\lambda_{\rm ij}^r$ are 
Mie potential parameters,
while $r_{\rm ij}$ is the distance between two CG beads.
Units are chosen for the length, $\sigma$, energy, $\epsilon$, mass, $m$, and
time $\tau$, which in real units would correspond to:
$\sigma = 0.43635$~nm, $\epsilon / k_B = 492$~K, $m=44.0521$~amu
and $\tau = \sigma{(m/{\epsilon})}^{0.5} = 1.4062$~ps. 
All simulations are carried out in the NVT ensemble by using the
Nos\'e--Hoover thermostat as implemented in the LAMMPS
package \cite{LAMMPS} with an integration time-step 
$\delta t = 0.005~\tau$. Our simulations took place
at room temperature ($T=25^{~\circ}$C), which in the
simulation units corresponds to $T=0.6057~\epsilon/k_B$.
Finally, a universal cutoff for all 
nonbonded (Mie) interactions is set to $r_c = 4.583~\sigma$.

\begin{figure}[bt!]
\centering
\includegraphics[width=\columnwidth]{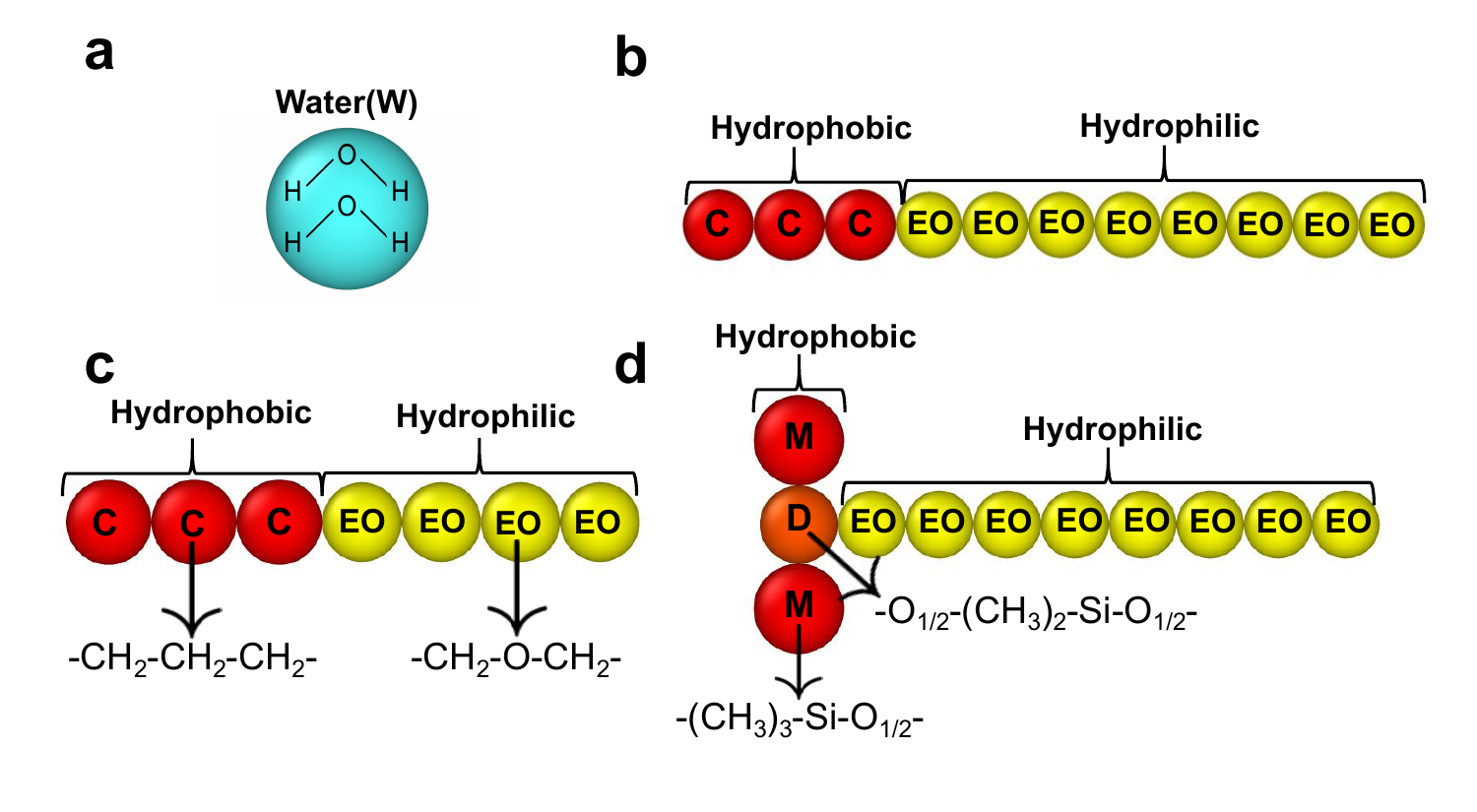}
\caption{\label{fig:2} Coarse-grained representation 
of water and surfactant molecules. Surfactant's 
hydrophobic beads are shown in red, while 
the hydrophilic parts of the surfactant are in
yellow. A cyan bead corresponds to two
water molecules (a). C10E8 (b), C10E4 (c),
and Silwet-L77 (d) CG surfactant models.
}
\end{figure}

We have considered surfactants
of type CiEj, such as C10E8 and C10E4 (Figs~\ref{fig:2}b, c) 
and a trisiloxane surfactant.\cite{Theodorakis2014,Theodorakis2015Langmuir, Theodorakis2019,Theodorakis2015modelling,Ritacco2010b,Ivanova2010,Ivanova2012,Ivanova2009} 
In the case of CiEj surfactants, a hydrophobic alkane
CG `C' bead represents a $\rm -CH_2-CH_2-CH_2-$ 
group of atoms,
while a hydrophilic CG `EO' bead represents an
oxyethylene group $\rm -CH_2-O-CH_2$.
Silwet-L77 (Fig.~\ref{fig:2}d)
is also considered as a trisiloxane surfactant with
the same number of beads as in the case of 
C10E8, but different hydrophobic chemical
units and architecture, 
where M type beads represent
a chemical group $\rm(-CH_3)_3-Si-O_{1/2}$ and 
D type $\rm O_{1/2}-(CH_3)_2-Si-O_{1/2}$.
Finally, a water CG `W' bead corresponds 
to two water molecules (Fig.~\ref{fig:2}a).
The nonbonded interaction parameters between
the above chemical groups, which can be
used in practice to simulate a wide range of surfactants with
different molecular architecture 
are reported in Table~\ref{tbl:table1},
while the mass of each CG bead is documented 
in Table \ref{tbl:table2}.

\begin{table}[bt!]
\small
    \caption{\ Summary of Mie interaction parameters (Eq.~\ref{equation_mie}). $\lambda_{\rm ij}^{\rm a} = 6$. }
    \label{tbl:table1}
    \begin{tabular*}{0.48\textwidth}{@{\extracolsep{\fill}}lccr}
    \hline
    \textrm{i--j}&
    \textrm{$\sigma_{\rm ij}~[\sigma]$}&\textrm{$\epsilon_{\rm ij}~[\epsilon/k_B]$}&\textrm{$\lambda_{\rm ij}^{\rm r}$} \\
    \hline
    W--W & 0.8584 & 0.8129 & 8.00\\
    W--C & 0.9292 & 0.5081 & 10.75 \\
    W--EO & 0.8946 & 0.9756 & 11.94 \\ 
    W--M & 1.0491 & 0.8132 & 13.72\\
    W--D & 0.9643 & 0.6311 & 10.38\\
    C--C & 1.0000 & 0.7000 & 15.00 \\
    C--EO & 0.9653 & 0.7154 & 16.86\\
    M--M & 1.2398 & 0.8998 & 26.00\\
    M--D & 1.1550 &0.7114 & 18.83 \\
    M--EO & 1.0853 & 0.8262 & 22.18 \\
    D--D & 1.0702 & 0.5081 & 13.90\\
    D--EO & 1.0004 & 0.6355 & 16.21 \\
    EO--EO & 0.9307 & 0.8067 & 19.00\\
        \hline
    \end{tabular*}
\end{table}
    
\begin{table}[bt!]
\small
    \caption{Mass of the CG beads.}
    \label{tbl:table2}
    \begin{tabular*}{0.48\textwidth}{@{\extracolsep{\fill}}cc}
    \hline
    \textrm{Bead Type}&
    \textrm{Mass~[m]} \\ 
    \hline
    W & 0.8179 \\
    C & 0.9552  \\
    EO & 1.0000\\ 
    M & 1.8588\\
    D & 1.6833\\
        \hline
    \end{tabular*}
\end{table}

To tether beads together in the case of surfactant
chains, a bond potential is required, which
in the case of this model is harmonic, \textit{i.e.}, 
\begin{equation}
\label{equation_bonded1}
    V(r_{\rm ij}) = 0.5k(r_{\rm ij}-\sigma_{\rm ij})^2
\end{equation}
where the harmonic constant $k = 295.33$~$\epsilon/\sigma^2$.
Moreover, EO  beads experience a harmonic angle potential,
\begin{equation}
\label{equation_bonded2}
    V_\theta(\theta_{\rm ijk}) = 0.5k_\theta(\theta_{\rm ijk}- \theta_0)^2,
\end{equation}
where $\theta_{\rm ijk}$ is the angle between consecutive
beads i, j and k (here, i, j, k indicate the order of EO beads instead of bead type), 
$k_\theta = 4.32$~$\epsilon/$rad$^2$, 
and $\theta_0 = 2.75$~rad is the equilibrium angle.
Further discussion on the model can be found in previous 
studies. \cite{Theodorakis2015Langmuir,Theodorakis2015modelling,Theodorakis2019}


To prepare the initial configuration of each system, individual
droplets were first equilibrated in the NVT ensemble. 
The total number of beads in the simulations was $10^5$ per 
initial droplet, with approximately
5\% evaporation into the gas. 
Droplet diameters were $\sim53\sigma$,
which is about 23~nm, similar to that
of several previous studies.\cite{perumanath2019droplet,arbabi2023b}
Careful consideration was given not only
to observing the energy of the system, but,
also, making sure that the distribution of surfactant clusters
has reached a dynamic equilibrium and that each
of them was able to diffuse a distance many times its size.
After equilibration of the individual droplets, 
the volume of the simulation
box was doubled and the two droplets (and the 
surrounding gas) were placed
next to each other as shown in Fig.~\ref{fig:1}a.
In this way,  roughly the same thermodynamic conditions are maintained
and further evaporation that would 
reduce the number of water molecules of the droplets is avoided.
The final size of the simulation box is also chosen large enough
to avoid the interaction of mirror images
of the droplets over the periodic boundary conditions.
Figure~\ref{fig:1} illustrates typical snapshots
at different stages during coalescence for 
cases below and above CAC.
For our study, we have considered a range
of different surfactant concentrations up to about 6$\times$CAC,
which cover the relevant span of phenomena. 
A summary of the mean values of various properties for
our systems is given in Table~\ref{tbl:table3}.
We can see that increasing surfactant concentration
slightly increases the size of
the droplet. Also, note that the CAC in the case
of Silwet-L77 in terms of wt\% is almost double 
that of C10E4 and C10E8 surfactants. 

\begin{table}[bt!]
\small
  \caption{\ Properties of individual droplets (equilibrium) }
  \label{tbl:table3}
  \begin{tabular*}{0.48\textwidth}{@{\extracolsep{\fill}}cccc}
    \hline
    Concentration ($wt\%$) & Diameter ($\sigma$)& Water Beads* &  \# Molecules\\
    \hline
    \textbf{C10E4}\\
    6.25 & 53.08 & 90466.09 & 714  \\
    12.37 & 53.02 & 85496.45 & 1429  \\
    24.18 & 53.62 & 75572.27 & 2857  \\
    35.48 & 54.14 & 65746.73 & 4286 \\
    46.02 & 54.63 & 55966.55 & 5714 \\
    CAC $\approx$ 7.5~wt\%\\
    \hline 
    \textbf{C10E8}\\
    6.25 & 52.74 & 90519.09 & 455  \\
    12.37 & 52.98 & 85500.09 & 909  \\
    24.18 & 53.40 & 75722.36 & 1818  \\
    35.48 & 53.91 & 65862.36 & 2727 \\
    46.02 & 54.33 & 56488.91 & 3636 \\
    CAC $\approx$ 7.5~wt\%\\
    \hline 
    \textbf{Silwet-L77}\\
    7.6 & 52.62 & 90438.45 & 455  \\
    14.8 & 53.01 & 85574.64 & 909  \\
    28.2 & 53.89 & 75786.36 & 1818  \\
    40.3 & 54.78 & 66312.82 & 2727 \\
    51.2 & 55.71 & 56787.91 & 3636 \\
    CAC $\approx$ 16.23~wt\%\\
    \hline
  \end{tabular*}
  * Indicates the average number of water beads.
\end{table}

\begin{figure}[bt!]
\centering
\includegraphics[width=\columnwidth]{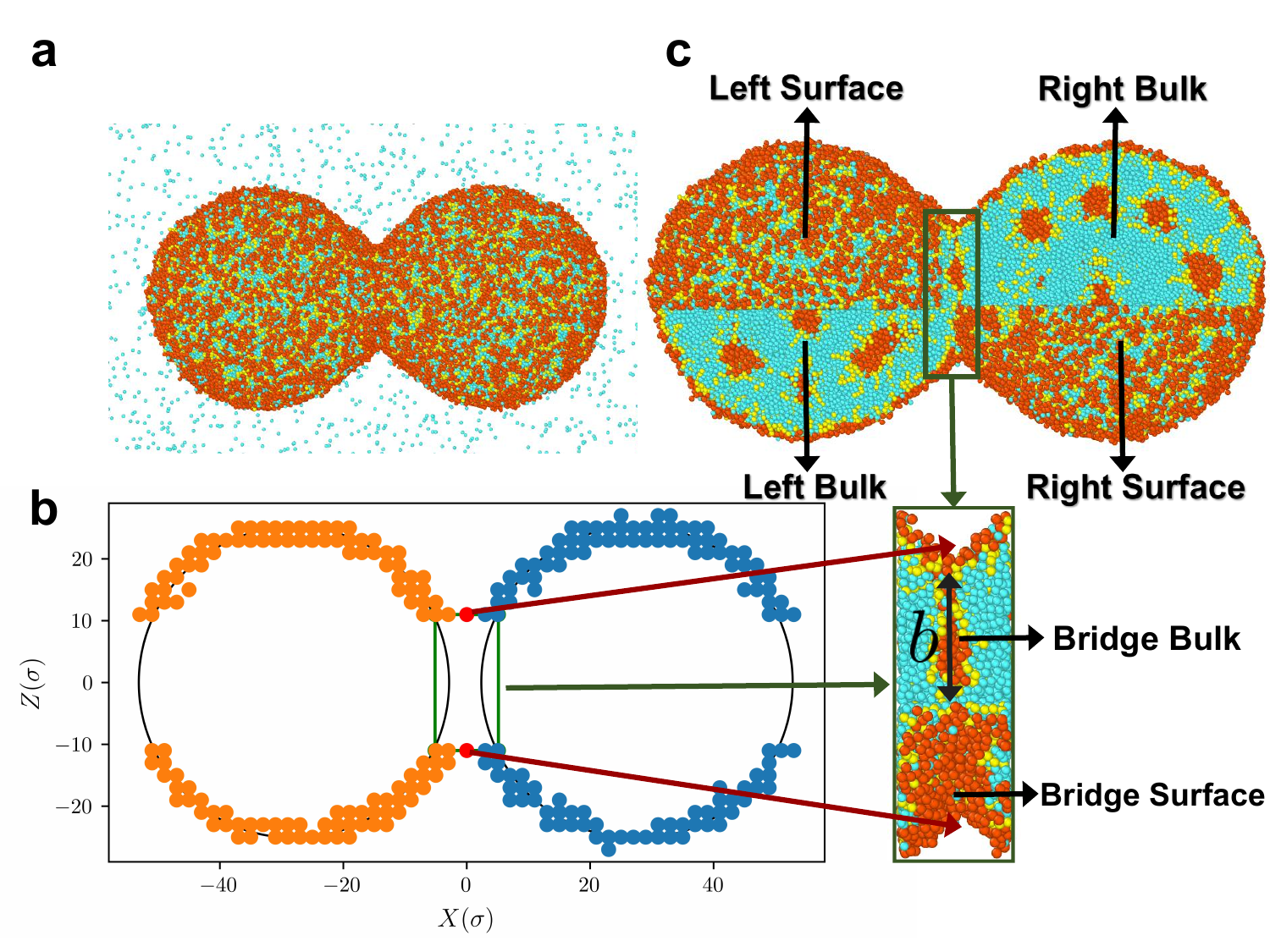}
\caption{\label{fig:3}  Bridge analysis and 
domains for analysing the mass transport mechanism.
Six different regions for the position of the 
molecules within the droplets are considered,
namely, bridge bulk and surface, left and right surface
and bulk. (a) A snapshot obtained by the MD simulation
with a clearly developed bridge length.
(b) Analysis of an $X-Z$ projection for identifying
the bridge based on the method described in Sec.~\ref{model}.
(c) Different regions of the droplets considered for
analysing the mass transport mechanism of surfactants
between these regions (see ESI for further details).
}
\end{figure}

To obtain reliable estimates of the bridge
growth dynamics and the mass transport mechanism,
snapshots of the system are made frequently
(every 250 MD time steps) for
the initial $4 \times 10^5$ MD time steps,
and cluster analysis is performed to identify
the beads belonging to the liquid phase (droplets),
which is used for our properties analysis.
The bridge region is chosen to be a slab with 
a width in the $X$ (approach) direction 
that is calculated for each configuration. 
In particular, the left and right limits of the slab are
determined by analysing the grid points
on the $X-Z$ plane at position $Y=0$ 
according to the
procedure shown in Fig.~\ref{fig:3}b. 
We fit a circle around each droplet
and note the surface grid positions 
at the central $X=0$ 
position, 
shown by the red points in Fig.~\ref{fig:3}b.
Horizontal lines are drawn in the $X$ direction
passing through these red points to touch the fitted circles,
thus defining the rectangle in green. 
The vertical sides of the rectangle give the limits 
of the bridge slab in the $X$ direction, and its width.
All molecules with centres having $X$ coordinates inside these 
limits are labelled as belonging to the bridge in a given 
snapshot. On the other hand, the bridge radius $b$ 
shown in  Figs~\ref{fig:1}c and \ref{fig:3} is calculated 
using the distances between extrema of 
the positions of the beads belonging to the grids 
located at $X=0$.
That is this distance is
first calculated separately for the $Z$ coordinate
to give a distance $2b_Z$,
and then for the $Y$ coordinate to give $2b_Y$. 
The final bridge radius estimate is then 
given by $b=(b_{Z}+b_{Y})/2$.

We also describe the mass transport mechanism
by tracking each surfactant molecule during 
the coalescence and then identifying the probability
of adsorption of each molecule to different
regions in the liquid phase, namely 
the left and right LG surfaces and the bulk,
and the surface and interior of the bridge (Fig.~\ref{fig:3}b, c).
Further discussion and details on the calculation
of probabilities related to the mass transport
mechanism can be found in the Electronic Supplementary 
Information (ESI). Finally, we have calculated
the density profiles of the water and surfactant
molecules during coalescence, the flow field
at different times, as well as the approach
distance and velocity of approach of the
droplets and their asphericity, 
for which further details are
discussed in the ESI.

\begin{figure}[bt!]
\centering
\includegraphics[width=\columnwidth]{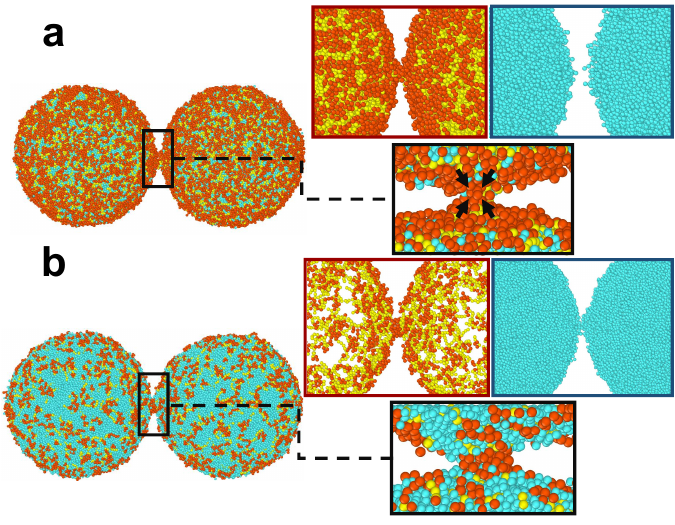}
\caption{\label{fig:4} Pinching of droplets with
C10E4 surfactant in the TR for concentration above (a, 35.4~wt\%) and 
below (b, 6.25~wt\%) CAC. For each case, a magnified view
of the initial bridge formation is shown,
as well as the distribution of surfactant (hydrophobic
beads in red and hydrophilic in yellow colour) and
water molecules (cyan colour). 
(a) $t=t_{\rm c}+8.75~\tau$,
(b) $t=t_{\rm c}+8.75~\tau$,
}
\end{figure}

\begin{figure}[bt!]
\centering
\includegraphics[width=\columnwidth]{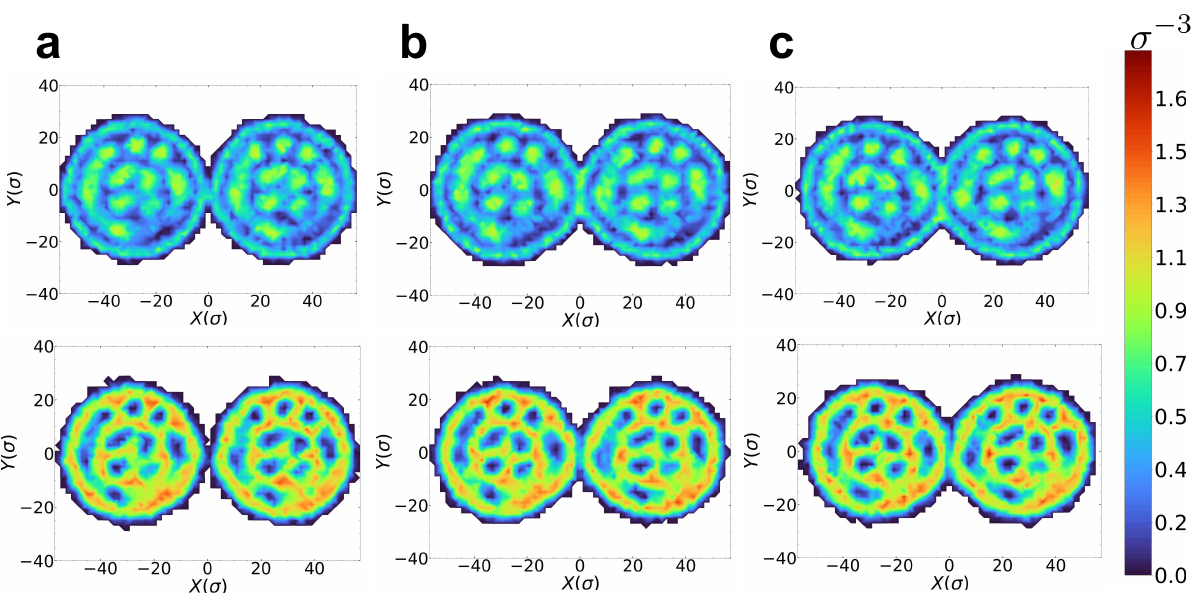}
\caption{\label{fig:5} Density profiles of surfactant beads,
C10E4 (46.2~wt\%)  (upper row) and water beads (lower row)
on the mid plane (width of $6~\sigma$) 
at different stages after the time $t_{\rm c}$ at which
a permanent contact between the droplets is established
(a) $t=t_{\rm c}~\tau$,
(b) $t=t_{\rm c}+71.25~\tau$,
(c) $t=t_{\rm c}+98.75~\tau$.
}
\end{figure}

\section{Results and Discussion}
\label{results}

\begin{figure*}[bt!]
\includegraphics[width=1.0\textwidth]{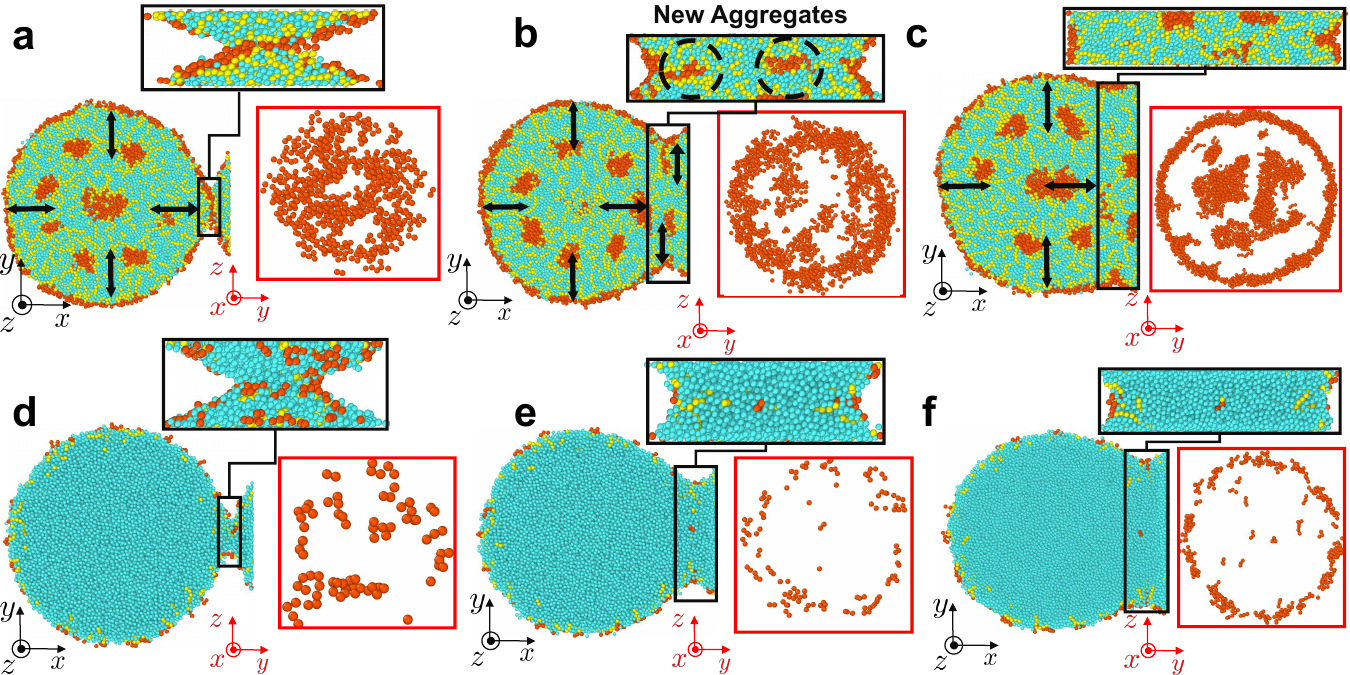}
\caption{\label{fig:6} Mass transport mechanism of 
surfactant (C10E8) during the coalescence process, 
for concentrations above (upper panel) and below (lower panel) CAC. 
The \blu{size} of the arrows reflects the probabilities
associated with surfactant transport to the different droplet areas 
(see Table~S4 of the ESI for further details). 
Above the CAC (a--c, 46.2~wt\%) snapshots were obtained 
at times (a)\,$t_{\rm c}+22.5~\tau$, (b)\,$t_{\rm c}+172.5~\tau$, 
(c)\,$t_{\rm c}+480.0~\tau$ while below the CAC (d--e, 6.2~wt\%) times
shown are (d)\,$t_{\rm c}+25.0~\tau$, (e)\,$t_{\rm c}+66.3~\tau$,
(f)\,$t_{\rm c}+116.3~\tau$,
soon after the end of the thermal regime (a,d), the development of the bridge and 
the formation of new aggregates (b) or surfactant monomers
remaining in the bridge region (e), and the full development
of the bridge (c, f). Magnified views of the bridge region and 
its cross-section (showing only surfactant hydrophobic beads in
the bridge region, red)
are attached above and to the right of the snapshots, respectively.
Figure~S4 in the ESI illustrates results for the C10E4 surfactant.
}
\end{figure*}

\subsection{Coalescence mechanism}
The initial stage of droplets' coalescence
manifests itself via their pinching as
illustrated by the snapshots of Fig.~\ref{fig:4}
and movies in the ESI.
In this early stage of the phenomenon, 
a previous study on coalescence with 
pure water droplets has shown via all-atom MD
that multiple bridges form on the surface
of the droplets and the 
overall radius of the affected region grows linearly
over time.\cite{perumanath2019droplet}
Here, we rather see that the droplet
pinching initially involves only the
surfactant molecules at the droplets' surfaces,
particularly when the surfactant concentration
is higher. Here and in those previous results the
size of the droplets are about the same.
Fig.~\ref{fig:4} illustrates the single
pinching of the droplets based on our model,
whereby hydrophobic parts of the
surfactants come together forming aggregates,
while water molecules remain further from 
the pinching point (Fig.~\ref{fig:4}).
By inspecting the density profile of
the droplets during coalescence and considering
a cross-section on the $X-Y$ plane
passing through the centre of mass
of the droplets for high-concentration
droplets, we can better highlight the
aggregates at the cross-section (Fig.~\ref{fig:5}). 
One can observe that the water density between
the droplets is negligible (the grid used for the calculation 
of the density profile does not resolve the bead size),
in contrast to the density of surfactant. Therefore,
the latter is solely responsible for the droplet
pinching. At later times and when the bridge
has developed past the very initial 
pinching stage, water molecules become part of
the bridge (Fig.~\ref{fig:5}b). 
However, the bridge region is still dominated by the presence of surfactant,
with surfactant density values similar to those inside aggregates.
It is notable that surfactant continues to
be present in the bridge in significant amounts even at later stages
of coalescence, when the bridge has been 
almost fully developed (Fig.~\ref{fig:5}c).
The reasons for this will become more
apparent when further details on the
surfactant mass transport are unveiled.

After the initial droplet pinching (Fig~\ref{fig:4}),
the contact between the droplet persists and
the growth of the bridge, which manifests by
the large change in curvature at the surface,
takes place as shown
in Fig.~\ref{fig:6}. Moreover, we analysed the
surfactant mass transport mechanism during coalescence,
and the main adsorption processes of 
surfactant were monitored. 
These processes are described by the probabilities
of surfactant remaining at a particular region or
moving between the different regions shown in 
Fig.~\ref{fig:3}c during coalescence.
These are documented in the ESI for all surfactant types
and concentrations considered in our study. 
On the basis of these probabilities we have
indicated by arrows the
main surfactant movements (Fig.~\ref{fig:6}). 
In particular, the concentration of surfactant
is high at the initial contact of
the droplets due to the preexisting surfactant at the 
droplet surfaces (Fig.~\ref{fig:6}a, d), which becomes
trapped in between the droplets.
As the bridge gradually grows, we observe that
most of the surfactant, which was initially on
the surfaces of the droplets, preferably moves
towards the surface of the bridge, which
is energetically more favourable.
However, the bridge surface has limited 
place for accommodating all surfactant from 
the initial LG surfaces of the droplets,
since in this area between the droplets
the surface excess concentration is doubled
at the initial approach.
As a result, some surfactant remains
in the bulk and forms new aggregates, which
is most clearly seen in the
case of higher surfactant concentrations 
(Fig.~\ref{fig:6}b, c and Figs.~S1--S3 and movie in
the ESI). In contrast, when the concentration is lower than CAC,
the bridge surface is able to accommodate
surfactant molecules that existed on the
LG surfaces before the coalescence
(see Tables~S1--S3 for detailed surfactant counts). 

\begin{figure}[bt!]
\includegraphics[width=1.0\columnwidth]{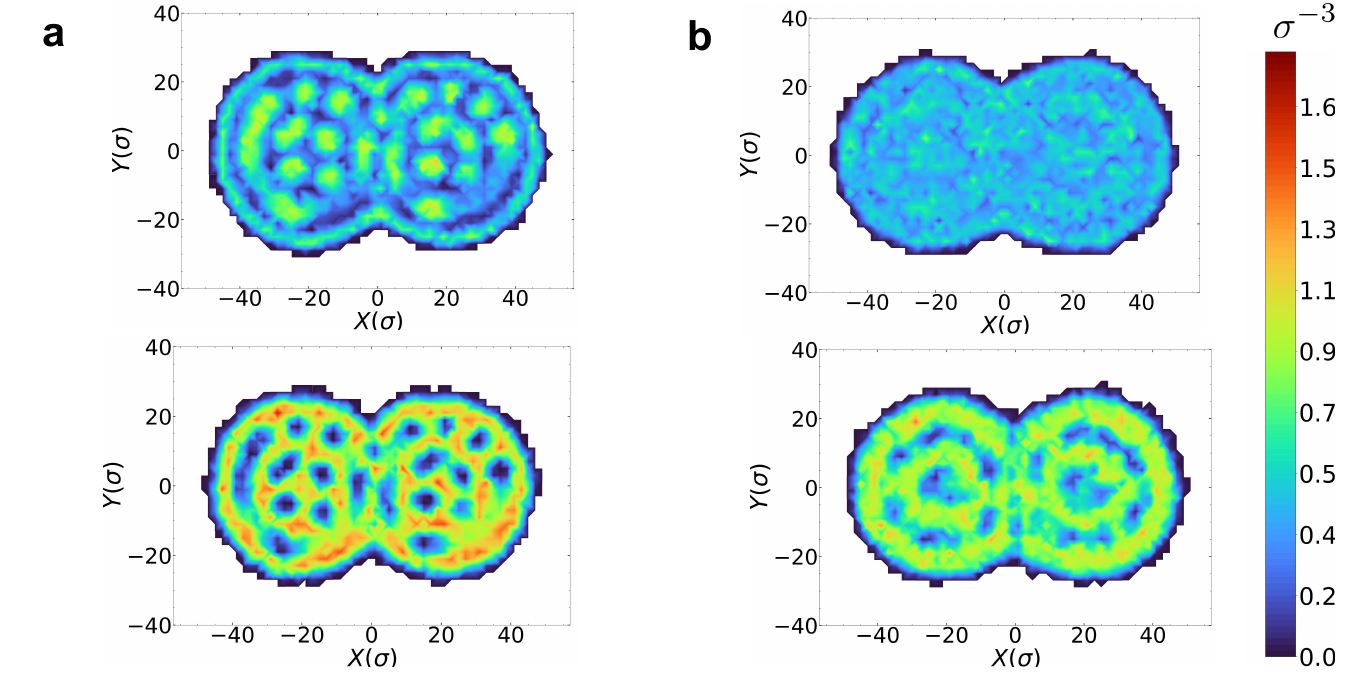}
\caption{\label{fig:7} Density profiles  of a $X-Y$
cross-section ($X$ and $Y$ are the coordinates of
the centre point of each grid voxel)
passing through the centre of mass
of the droplets with (a) C10E4 
(46.2~wt\%) \,$t=t_{\rm c}+288.75~\tau$
and (b) Silwet-L77 (51.2~wt\%), $t=t_{\rm c}+437.50~\tau$. 
Upper panels show the number density of the surfactant beads,
lower panels of the water.
}
\end{figure}

We observe that most of the
surfactant activity takes place in
the bridge area and that the dominant
process is transport towards the bridge surface. 
We do not see any clear evidence of Marangoni flow in the studied system, 
possibly due to there not being enough space or time for it to develop.
By analysing our data, we have verified that the qualitative features 
of the mass transport mechanism are
independent of the surfactant type or concentration, 
while the largest quantitative differences were seen
in the case of Silwet-L77. In particular, Silwet-L77 
tends to form a larger number of aggregates but their
overall density is lower than that of the C10E4 and C10E8
surfactants, which also implies that higher quantities of water molecules
are found among surfactant in the case
of droplets with Silwet-L77. To illustrate
these effects, the density profiles of droplets
with either C10E4 or Silwet-L77 with high surfactant concentration are plotted
in Fig.~\ref{fig:7}. Finally, in terms of the mass transport (Table S4 in
the ESI contains details), we find that Silwet-L77 have a higher tendency to
move towards the surface of the bridge in comparison with C10E4 and C10E8 surfactants,
which is in line with previous observations in systems with surfactant-laden 
droplets.\cite{Theodorakis2019} In turn, this results in the formation of
fewer aggregates in the bridge in the case of droplets with Silwet-L77 surfactant.

\begin{figure}[bt!]
\includegraphics[width=1.0\columnwidth]{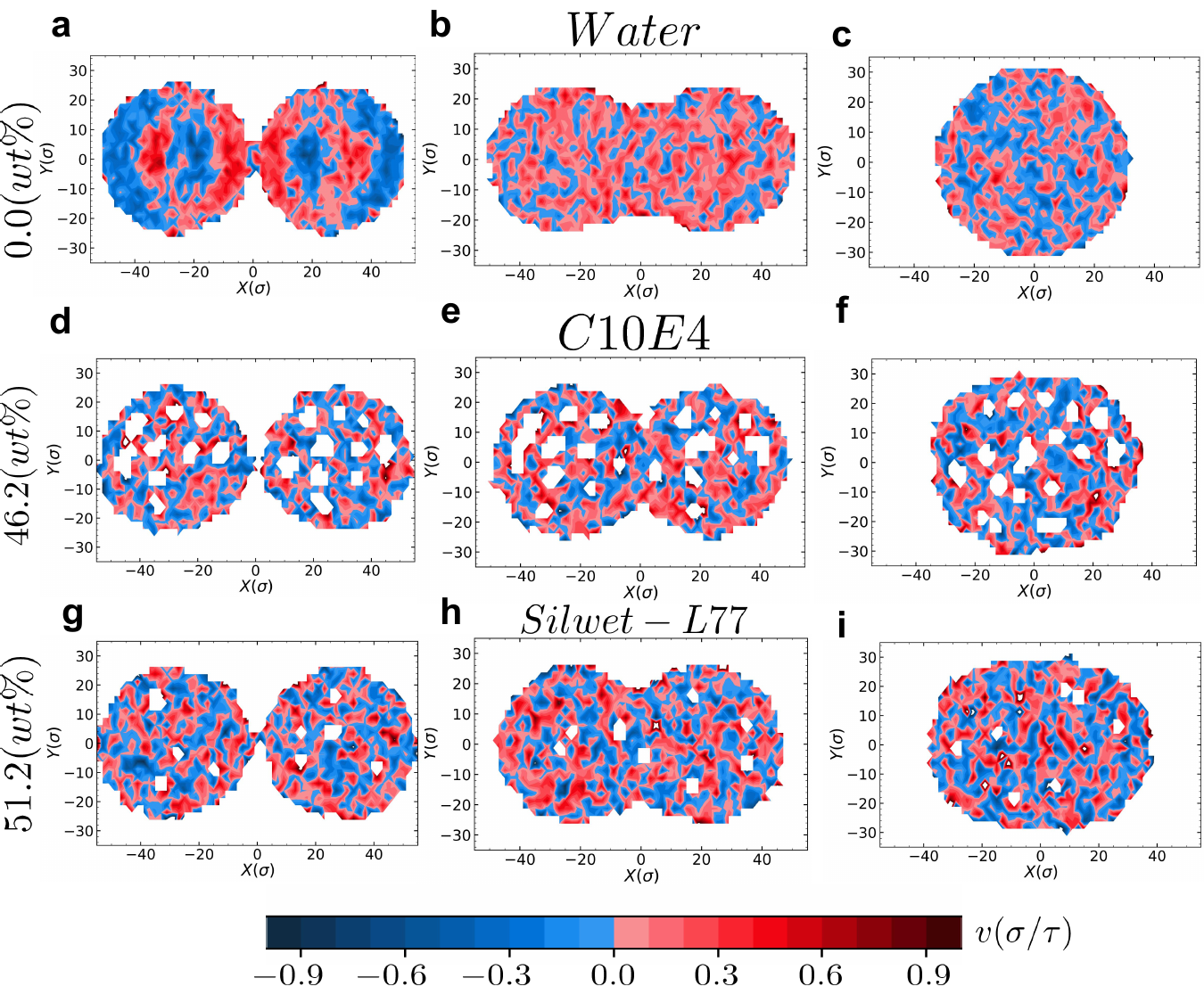}
\caption{\label{fig:8} Flow field of water inside the 
droplets for the case without surfactant (a--c),
with C10E4 (46.2~wt\%, d--f), and 
with Silwet-L77 (51.2~wt\%, g--i) surfactant. 
Cross-sections through the centre of mass are shown. 
Red reflects the intensity of flow motion (only water)
toward the bridge $X$ component of velocity $\varv$, 
blue away from the bridge. The averaged data of red and
blue grids indicating the total flow towards the bridge
and away of it in the droplets and in the bridge 
region are reported in the ESI and discussed further
in the main text. Note that the white
space between the water areas (e.g. in the bridge) includes
surfactant aggregates and surfactant on the surface.
The snapshots were obtained at times (a, d, g) $t_{\rm c}$
(b) $t_{\rm c}+90.75~\tau$, (c) $t_{\rm c}+275.25~\tau$,
(e) $t_{\rm c}+183.75~\tau$, (f) $t_{\rm c}+2446~\tau$,
(h) $t_{\rm c}+425~\tau$, and (i) $t_{\rm c}+2437.5~\tau$.
}
\end{figure}

\subsection{Water flow}
In a previous study,\cite{arbabi2023b} we have shown that
the coalescence of water droplets is characterised by
intense internal flow variations during the thermal regime,
which attenuate with the increase of surfactant
concentration.
Figure~\ref{fig:8} illustrates the flow of water
in the case of water droplets and those with
surfactant after the initial
thermal regime and at the initial stages of the
power-law (IR) regime. Firstly, as expected, the data suggest that there
is an increased water flow toward the bridge (red colour)
throughout the coalescence process. This is particularly visible in the bulk in the IR phase, as  
can be better seen in data included in Tables~S6--S8 in the ESI. These 
are separately averaged for the red and the blue
grids and concern
the pure water and surfactant-laden droplets at high concentration. Moreover, 
surfactant attenuates the free flow of water molecules
toward the bridge, because it
reduces the surface tension at the droplets' LG
surface and moreover forms aggregates in the bulk that
hinder the flow directly. The formation of aggregates
at the pinching point is also seen, which manifests
by the empty spots in the flow field of
Fig.~\ref{fig:8}. These observations are valid
for all of the different surfactants studied here
and for the whole range of concentrations above CAC.

\begin{figure}[bt!]
\centering
\includegraphics[width=\columnwidth]{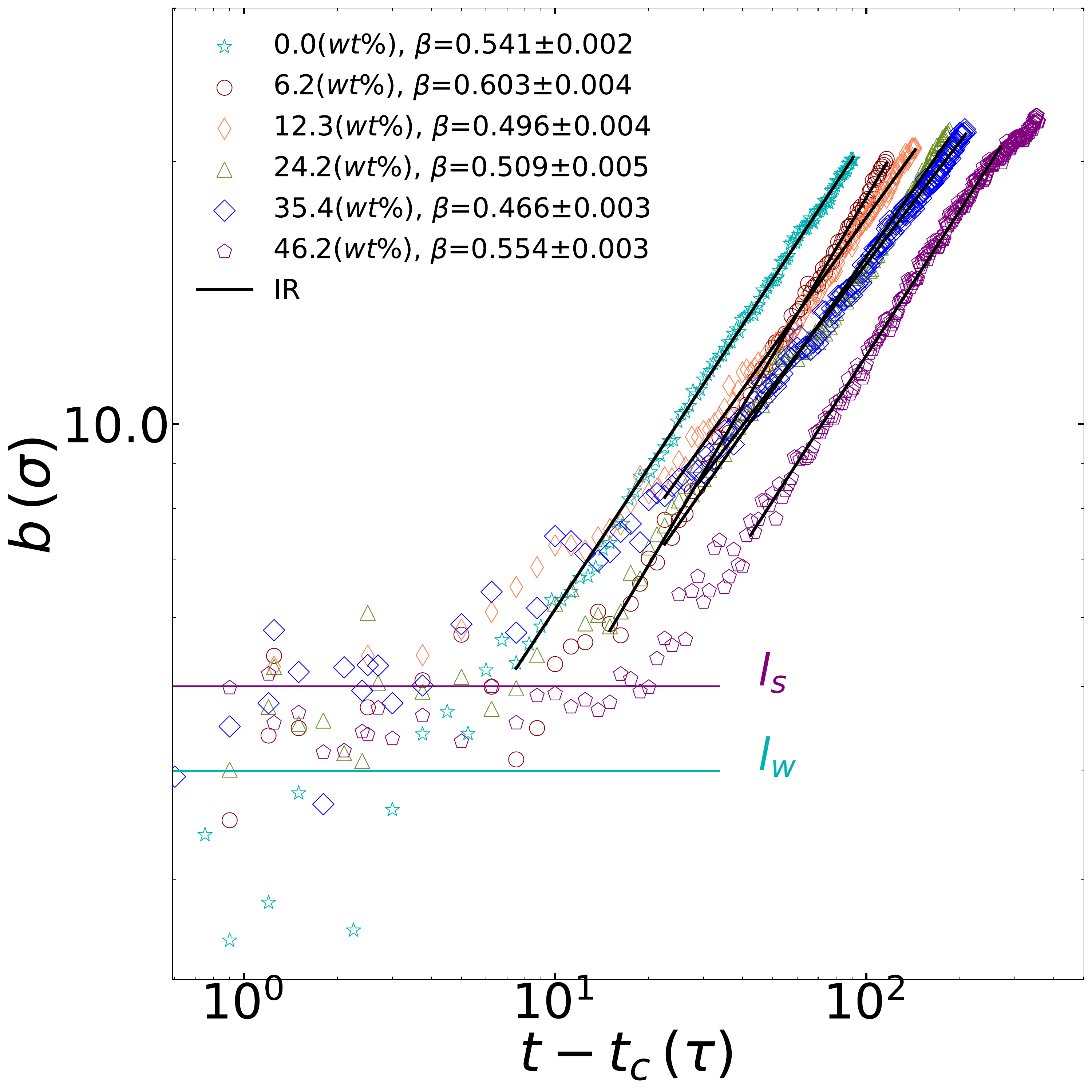}
\caption{\label{fig:9} Bridge length $b$ as a function
of time from the first contact of the droplets at
$t_c$ for different concentrations of C10E4
surfactant, as indicated. Power-law fits $\sim t^\beta$ 
are shown, labeled ``IR''. $l_s$ is the thermal length in the case
of surfactant-laden droplets above CAC, while
$l_w$ is the thermal length in the case of pure water
droplets.
}
\end{figure}

\subsection{Bridge Growth}
\begin{figure}[bt!]
\centering
\includegraphics[width=\columnwidth]{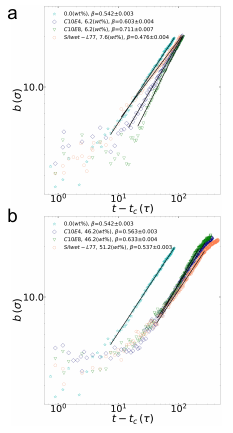}
\caption{\label{fig:10} Comparison of bridge radius growth $b(t)$ for different surfactants,
below (a) and significantly above (b) CAC.
Power-law fits $\sim t^\beta$ 
are shown, for times counted from the first contact of 
the droplets at
$t_c$.
}
\end{figure}

The growth rate of the bridge radius, $b$, 
is a key parameter that characterises
the dynamics of the coalescence process. 
From a fluid dynamics perspective, the VR
and IR have mainly been proposed, where
the former manifests itself through a linear dependence on
time, $b\sim t$, while the latter is described by a 
slower power-law dependence, namely $b \sim t^\beta$\cite{paulsen2012inexorable,paulsen2013approach,frenkel1945viscous,eggers1999coalescence}
with expected $\beta\approx\tfrac{1}{2}$.
In addition, all-atom molecular dynamics have 
identified a thermal regime at the very early
stage of coalescence, which persists over the length scale
$l_{T}$ as mentioned in Sec.\ \ref{intro}.\cite{perumanath2019droplet}
Well beyond this length, the hydrodynamic description is considered 
valid. In view of the importance of $b$ in 
describing the dynamics of coalescence and the
various scenarios discussed thus far in the literature,
we have embarked here on investigating the time
evolution of $b$ for a range of
different surfactants and concentrations. 
Figure~\ref{fig:9} presents results for droplets with
C10E4 surfactant at different concentrations, both
below and above the CAC. In Fig.~S6 in the ESI, results
for other surfactants and concentrations are shown.
Overall, our model captures two different 
coalescence regimes, namely an initial TR
and a subsequent power-law regime characterised
by power law exponents $\beta$ in the range $0.46-0.71$, 
mean $\beta=0.57$ close to those predicted by IR scaling. 
We observe different exponents for different surfactant
concentration, with more or less random variation from 
case to case, which may be due to finite size effects 
and different internal aggregate configurations.
Prefactors generally decrease with growing surfactant 
concentration, leading to a slowdown of growth.
Hence, as in the case of all-atom
water simulation,\cite{perumanath2019droplet} our CG model
is also confirming the existence of a TR and an inertial-like
power-law regime. The TR persists over a time
$\mathcal{O}(10~\tau)$ during which the 
length of the bridge is well described by the length 
$l_{T}$ both for the pure and the surfactant-laden droplets.
In Fig.~\ref{fig:10}, we compare the bridge growth for three 
different surfactants below and well above CAC to each other
and pure water. Differences in the surface
tension above CAC for the different surfactants 
are generally expected to be small, and our results 
for $b(t)$ also differ a little, 
which agrees with the expectation that bridge size 
in this regime is determined by the thermal 
length $l_{T}\propto\gamma^{-1/4}$.

Following the TR, we observe the transition to 
the power-law regime. The time for the transition to
occur is longer when droplets have a higher 
surfactant concentration. The power-law regime 
is characterised by exponents close and generally
above $0.5$. The mean over all cases is $\beta=0.57$ and standard deviation $0.06$, while the reference pure water case gives $\beta=0.541$. It is also clearly seen that 
adding surfactant will
lead to slower dynamics overall, regardless of exponent variability,  as well as a delay in the start of the bridge growth --- see ESI Fig.~S5.
We can also see a trend that C10E8 clearly has higher 
$\beta$ exponent than pure water and than C10E4
(Compare Fig.~S6 in the ESI with Fig.~\ref{fig:9}). Otherwise, 
the main variation lies in the reduction of the growth prefactor
as surfactant concentration grows, as can be seen 
by the progressive shifting of the evolution 
to the right in the logarithmic plots of Figs.~\ref{fig:9},
 \ref{fig:10}, and ESI Fig.~S6. This trend is broadly similar 
for all surfactants that we studied, 
with the addition that Silwet-L77 displayed 
a significantly slower growth in all respects, and in particular 
lower maximum velocity (see ESI, Figs.~S7--S8).

\begin{figure}[bt!]
\centering
\includegraphics[width=\columnwidth]{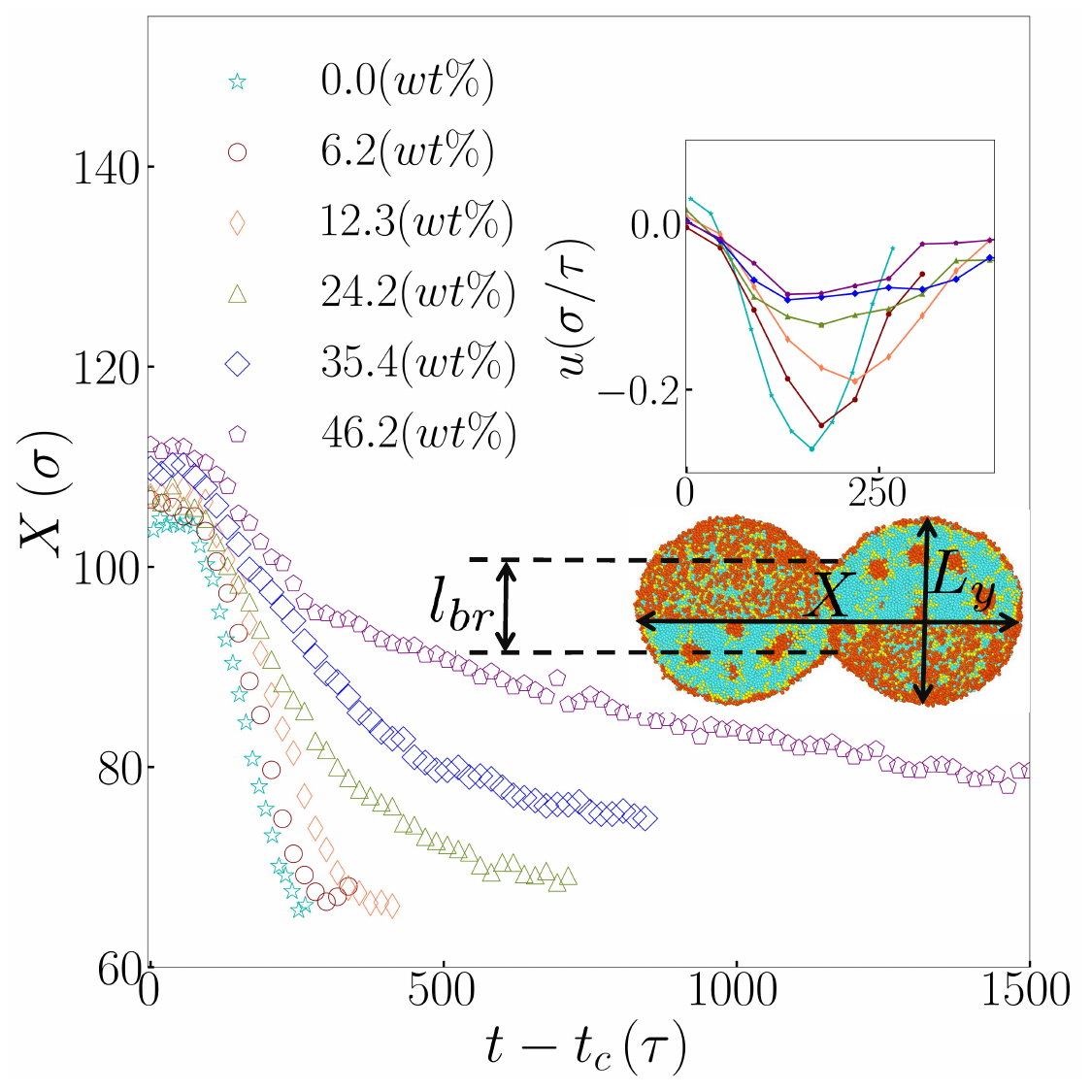}
\caption{\label{fig:11} System length, $X$, in the coalescence direction 
for different concentrations of C10E4. The inset
shows the instantaneous velocity of approach $u = dX/dt$.
The maximum speed occurs at droplet-to-bridge width ratios of 
$L_y/l_{br}$=1.22 ($0.0$~wt\%), 1.22 ($6.2$~wt\%), 1.30 ($12.3$~wt\%)),
1.40 ($24.2$~wt\%), 1.80 ($35.4$~wt\%), 1.80 ($46.2$~wt\%).
}
\end{figure}

\subsection{Velocity of approach and asphericity}
A complementary measure of the coalescence dynamics is obtained by 
monitoring the total length of the droplets in the $X$
direction, whence the velocity of approach $u=\dot{X}$ can be 
obtained.\cite{Verdier2002,Arbabi2023} Data for droplets with
C10E4 surfactant are presented in Fig.~\ref{fig:11},
and show that the fastest approach takes place in the middle of the 
process at a time about 150~$\tau$, which
is similar for different surfactants.
Although the exponents $\beta$ of the bridge radius
are roughly similar for all cases, we observe
that the droplets with surfactant consistently exhibit a smaller
maximum speed. Moreover, the process is shorter in time in the case
of pure water droplets.  
A different behaviour appears well above CAC (e.g. four times above), 
when the process becomes significantly slower, sometimes rather suddenly.
A conjecture is that this is due to increased 
internal rigidity of the aggregates present in the bulk.
Monitoring the ratio of droplet radius to bridge radius
$L_y/l_{br}\approx R_0/b$ one finds that
maximum coalescence velocity occurs for very broad bridges 
$L_y/l_{br}\approx 1.2$ when the surfactant concentration 
is low but earlier and for much smaller bridge size
$L_y/l_{br}\approx 1.8$ when well above CAC. 
This would be consistent with the conjecture
that a major slowdown occurs once significant internal 
rearrangements, as necessitated by a broad bridge in high concentrations, 
become necessary. Further results for different surfactants and comparative
plots are shown in the ESI (Figs~S7 and S8).

Monitoring the asphericity
of the two coalescing droplets (Figs~S9 and S10 in
the ESI), as a complimentary measure to monitor
the lateral changes in the dimensions of 
the system, we can observe that this generally follows the
behaviour of the velocity of approach.
Pure water droplets complete the coalescing process
obtaining the final spherical shape
much faster than in the case of
droplets with surfactant. As with approach velocity and system size, 
we see a change in behaviour when surfactant
concentration increases, but the change is gradual
and does not occur at the CAC but well above.

\section{Conclusions}
\label{conclusions}
In this study, we have analysed various 
macroscopic and microscopic properties of droplets laden
with surfactant and unveiled the surfactant
mass transport mechanism during the process for
different surfactants and a range of 
concentrations. We have demonstrated that
the underlying mechanisms and flow patterns of coalescence are
universal, qualitatively and often quantitatively
independent of the type of surfactant.
Particular differences between Silwet-L77 and
CiEj surfactants have been noted in
the bridge dynamics for concentrations below the CAC. 
We have also observed that water
molecules are not part of the initial pinching process
of coalescence \blu{at larger (above CAC) surfactant concentration},
which is driven by 
surfactant aggregation at the droplet's surface.

Other features include the engulfment of part of
the initial contact surface inside the forming bridge, the
existence of an initial thermal regime, followed by a later
power-law growth of the bridge radius with power exponents
close to $\frac{1}{2}$, as expected in the inertial regime.
We saw no evidence of an intermediate viscous regime.
One possibility is that the droplets, despite being large
computationally, were still too small for the VR to emerge
out of the thermal regime before the IR is activated. 
Coalescence also universally becomes slower as surfactant 
concentration grows, and we see evidence of the appearance
of a further slowdown with different qualities for several 
times the critical concentration, using several different 
indicators. We conjecture that this is due to the appearance
of greater internal stiffness caused by closely-packed 
surfactant aggregates in the bulk. 
The range of concentrations studied here reflects
the need in practical applications where surfactant
concentration is usually above the 
CAC.\cite{gonccalves2022experimental}
We anticipate that our results demonstrate the
mechanisms of a fundamental process in nature
and technological applications, which remain
universal.

\FloatBarrier

\section*{Conflicts of interest}
There are no conflicts to declare

\section*{Acknowledgements}
This research has been supported by the 
National Science Centre, Poland, under
grant No.\ 2019/34/E/ST3/00232. 
We gratefully acknowledge Polish high-performance computing infrastructure PLGrid (HPC Centers: ACK Cyfronet AGH) for providing computer facilities and support within computational grant no. PLG/2022/015747.



\balance


\bibliography{rsc} 
\bibliographystyle{rsc} 

\end{document}